# Impact of Weight Loss on Brain Age: Improved Brain Health Following Bariatric Surgery

Yashar Zeighami[1,2,*], Mahsa Dadar[3], Justine Daoust[4], Mélissa Pelletier[4], Laurent Biertho[5], Léonie Bouvet-Bouchard[5], Stephanie Fulton[6], André Tchernof[4], Alain Dagher[1], Denis Richard[4], Alan Evans[1,2], Andréanne Michaud[4,*]

1: Montreal Neurological Institute, Department of Neurology and Neurosurgery, McGill University, Montreal, Canada

2: Ludmer Centre for Neuroinformatics and Mental Health, McGill University, Montreal, Canada

3: CERVO Brain Research Center, Centre intégré universitaire santé et services sociaux de la Capitale Nationale, Université Laval, Québec, Canada

4: Centre de recherche de l'Institut universitaire de cardiologie et de pneumologie de Québec, Université Laval, Québec, Canada

5: Département de chirurgie générale, Institut universitaire de cardiologie et de pneumologie de Québec, Université Laval, Québec, Canada

6: Centre de Recherche du CHUM, Department of Nutrition, Université de Montréal, Montreal Diabetes Research Center, Montreal, QC, Canada.

***Correspondenc*e:**
Andréanne Michaud, Dt.P., Ph.D.
E-mail: andreanne.michaud@fsaa.ulaval.ca

Yashar Zeighami Ph.D.
E-mail: yashar.zeighami@mcgill.ca

**Abstract**

Overweight and obese individuals tend to have increased brain age, reflecting poorer brain health likely due to grey and white matter atrophy related to obesity. However, it is unclear if older brain age associated with obesity can be reversed following weight loss and cardiometabolic health improvement. The aim of this study was to assess the impact of weight loss and cardiometabolic improvement following bariatric surgery on brain health, as measured by change in brain age estimated based on voxel-based morphometry (VBM) measurements. We used three distinct datasets to perform this study: 1) CamCAN dataset to train the brain age prediction model, 2) Human Connectome Project (HCP) dataset to investigate whether individuals with obesity have greater brain age than individuals with normal weight, and 3) pre-surgery, as well as 4, 12, and 24 month post-surgery data from participants (n=87, age: 44.0±9.2 years, BMI: 43.9±4.2 kg/m$^2$) who underwent a bariatric surgery to investigate whether weight loss and cardiometabolic improvement as a result of bariatric surgery lowers the brain age. As expected, our results from the HCP dataset showed a higher brain age for individuals with obesity compared to individuals with normal weight (T-value = 7.08, p-value < 0.0001). We also found significant improvement in brain health, indicated by a decrease of 2.9 and 5.6 years in adjusted delta age at 12 and 24 months following bariatric surgery compared to baseline (p-value < 0.0005 for both). While the overall effect seemed to be driven by a global change across all brain regions and not from a specific region, our exploratory analysis showed lower delta age in certain brain regions (mainly in somatomotor, visual, and ventral attention networks) at 24 months. This reduced age was also associated with post-surgery improvements in BMI, systolic/diastolic blood pressure, and HOMA-IR (T-value$_{BMI}$=3.59, T-value$_{SBP}$=4.26, T-value$_{DBP}$=3.67, T-value$_{HOMA-IR}$=2.81, all p-values < 0.05). In conclusion, these results suggest that obesity-related brain health abnormalities (as measured by delta age) might be reversed by bariatric surgery-induced weight loss and widespread improvements in cardiometabolic alterations.

**Keywords:** Brain age, weight loss, bariatric surgery, voxel-based morphometry

**Introduction**

The human brain experiences morphological changes across the adult lifespan that are generally associated with a decline in cognitive performance as well as other behavioural and motor symptoms. The inter-individual variability in age-associated brain changes has been related to clinical outcomes (e.g. cognitive impairment and dementia), and the predicted age based on such brain changes can be used as a measure of brain health[3]. More specifically, the difference between brain age predicted based on features derived from magnetic resonance images (named brain age hereafter) and chronological age is defined as delta age or brain age gap estimate[4]. This delta age provides a measure of whether an individual's brain appears older or younger than a normal age-matched brain[3,5].

Previous studies have demonstrated increases in brain age in various disorders, such as in individuals with mild cognitive impairment progressing to Alzheimer's dementia[6,7], schizophrenia[8,9], HIV[10], epilepsy[11], Down's syndrome[12], major depressive disorder[13], and diabetes[14]. Brain age can also reliably predict cognitive impairment[15,16], future cognitive decline and dementia[7], and mortality[17]. Moreover, brain age has been identified as a better predictor of cognitive impairment compared to chronological age[18].

A number of brain imaging studies have also reported that individuals who are overweight or obese tend to have increased brain age[1,2], reflecting poorer brain health likely due to grey and white matter atrophy related to obesity[19–22]. Interestingly, grey matter reductions associated with obesity are consistent with age-related grey matter atrophy patterns[23], highlighting the importance of obesity prevention in promoting healthy aging. There is an increasing number of studies suggesting that at least part of the structural grey and white matter abnormalities associated with obesity might be driven by abdominal obesity-related cardiometabolic alterations, such as inflammation, insulin resistance, dyslipidemia, and hypertension[24–27]. Intriguingly, recent discoveries from human studies reported that cardiometabolic risk factors, including high blood pressure, markers of liver and kidney dysfunctions, diabetes, dyslipidemia, history of stroke, body mass index (BMI), and smoking are associated with an older-appearing brain and accelerated brain aging[28–30].

It is unknown whether interventions targeting weight loss and cardiometabolic improvement can reduce brain age and improve brain health. Bariatric surgery represents an interesting approach to examine the impact of marked weight loss and cardiometabolic improvements on brain health in a longitudinal setting. However, studies linking bariatric surgery-induced weight loss to reduced brain age and improved brain health are lacking. Using our unique longitudinal dataset following individuals that went through significant weight loss and cardiometabolic improvement following bariatric surgery, we investigated this hypothesis. Specifically, we determined: 1) whether individuals with obesity have higher brain age than individuals with normal weight, 2) whether brain age decreases 4, 12, and 24 months post-bariatric surgery (comparing pre- and post-surgery) and 3) whether this reduced brain age is associated with improvement in cardiometabolic risk factors.

**Methods**

**Data**

Three distinct datasets were used to: 1) train the brain age prediction model, 2) investigate whether individuals with obesity have greater brain age than individuals with normal weight, and 3) investigate whether weight loss as a result of bariatric surgery reverses the effect of obesity and lowers the brain age.

*1) Training dataset*

Data used to train the brain age prediction model included participants with T1-weighted MRI data available from the second stage of the Cambridge Centre for Ageing and Neuroscience (CamCAN, https://www.cam-can.org/index.php?content=dataset) dataset, described in Shafto et al. (2014)[31] and Taylor et al. (2017)[32]. Participants were screened for neurological and psychiatric conditions and those with such underlying disorders were excluded from the study. T1-weighted MRIs were acquired on a 3T Siemens TIM Trio, with a 32-channel head-coil using a 3D magnetization-prepared rapid gradient echo (MPRAGE) sequence (TR = 2,250 ms, TE = 2.99 ms, TI = 900 ms, FA = 9 deg, field of view (FOV) = 256 × 240 × 192 mm, 1 mm$^3$ isotropic, GRAPPA = 2, TA = 4 min 32 s). For detailed acquisition parameters see: https://camcan-archive.mrc-cbu.cam.ac.uk/dataaccess/pdfs/CAMCAN700_MR_params.pdf.

*2) Independent dataset (HCP) to compare brain age between individuals with obesity and individuals with normal weight*

Data used to assess the predicted brain age difference between individuals with obesity and those with normal weight included an independent sample from the Human Connectome Project (HCP)[33,34]. All the participants from the HCP with a BMI higher than 35 kg/m$^2$ were included in this sample. These participants (n = 46) were individually matched (1:1) for age, sex, and ethnicity with a group of HCP individuals who had a normal body weight (n = 46). Other exclusion criteria included participants with missing information on age, sex, BMI, and ethnicity. T1-weighted 3D MPRAGE sequence with 0.7 mm$^3$ isotropic resolution images were acquired by the HCP investigators using a 3T MRI scanner (Siemens Skyra) equipped with a 32-channel head coil. The following parameters were used: 256 sagittal slices in a single slab, TR = 2400 ms, TE = 2.14 ms, FA = 8 deg, FOV = 224 mm, TI = 1000 ms, Echo Spacing = 7.6 ms, voxel size = 0.7 × 0.7 × 0.7 mm$^3$ [35].

*3) Bariatric surgery-induced weight loss assessment dataset*

Data used for the main analyses included 87 participants with severe obesity (22 men, 65 women; mean age at baseline = 44.0 ± 9.2 years; mean BMI at baseline = 43.9 ± 4.2 kg/m$^2$) who underwent bariatric surgery at the *Institut universitaire de cardiologie et de pneumologie de Québec* (IUCPQ). In total, 87 participants were included at baseline, 71 participants at 4 months post-surgery, 47 participants at 12 months post-surgery, and 34 participants at 24 months post-surgery (**Table 1**). Inclusion criteria were: 1) individuals with BMI ≥35 kg/m$^2$ who required surgery and who met the NIH Guidelines for bariatric surgery[36]; 2) age between 18 and 60 years. Exclusion criteria were the following: 1) any uncontrolled medical, surgical, neurological or psychiatric condition; 2) liver cirrhosis or albumin deficiency; 3) medication that may affect the central nervous system; 4) pregnancy; 5) substance or alcohol abuse; 6) previous gastric, oesophageal, brain or bariatric surgery; 7) gastro-intestinal inflammatory diseases or gastro-intestinal ulcers; 8) severe food allergy; and 9) contraindications to MRI (implanted medical device, metal fragment in body, or claustrophobia). The Research Ethics Committee of the *Centre de recherche de l'IUCPQ* approved the study. All participants provided written informed consent to participate in the study.

Most participants (n=46) underwent a laparoscopic sleeve gastrectomy (SG), a restrictive surgery consisting of 150 to 250 cm$^3$ vertical gastrectomy on a 34 French bougie starting 4 to 5 cm proximal to the pylorus[37]. Laparoscopic biliopancreatic derivation with duodenal switch (BPD-DS) was performed in 12 participants, which is a mixed-surgery combining restrictive and malabsorptive mechanisms by creating a 150 to 250 cm$^3$ vertical SG and duodeno-ileal anastomosis 100 cm from ileocecal valve[38]. Thirteen participants underwent a laparoscopic Roux-en-Y gastric bypass (RYGB) surgery in which proximal gastric pouch of 30-50 cm$^3$ is created and anastomosed to the proximal small bowel by bypassing the first 100 cm and bringing a 100 cm alimentary limb on the gastric pouch.

The study design has been described in detail in Michaud et al. 2020[21]. Briefly, prior to (approximately 2 months prior to surgery) as well as 4, 12, and 24 months post-surgery, participants underwent a physical examination (blood pressure and anthropometric measurements), fasting blood sample, and an MRI session. They were asked to fast for 12h before the MRI session (scheduled between 9:00 and 10:30 am). One hour before scanning, they consumed a standardized beverage meal (237 ml, Boost original, Nestle Health Science) for a 5 minutes period to control hunger level during the MRI session. Hip, waist, and neck circumferences were measured in centimeters and were conducted using standardized procedures. Body weight and body composition were obtained using a calibrated bioelectrical impedance scale (InBody520, Biospace, Los Angeles, CA, or Tanita DC-430U, Arlington Heights, IL) to calculate body mass index BMI (kg/m$^2$), percentage of excess weight loss (%EWL), and percentage of total weight loss (%TWL) as previously described in Michaud et al. 2020[21]. Plasma levels of cholesterol, high-density lipoproteins, low-density lipoproteins, triglycerides, glucose and insulin were measured at the *IUCPQ* biochemistry laboratory. The homeostatic model assessment insulin resistance (HOMA-IR) index was calculated with the following formula: (fasting insulin (pmol/L) x fasting glucose (mmol/L)) / (22.5 x 6).

T1-weighted three-dimensional (3D) turbo field echo images were acquired using a 3T whole-body MRI scanner (Philips, Ingenia, Philips Medical Systems) equipped with a 32-channel head coil at the *Centre de recherche de l'Institut universitaire de cardiologie et pneumologie de Québec*. The following parameters were used: 176 sagittal 1.0 mm slices, TR = 8.1 ms, TE = 3.7 ms, FOV = 240 × 240 mm$^2$, and voxel size = 1 × 1 × 1 mm$^3$.

**Table 1.** Characteristics of participants at baseline as well as 4, 12, and 24 months after bariatric surgery

| Variable | Baseline | 4 months | 12 months | 24 months | *p* value |
|---|---|---|---|---|---|
| *N* | 87 | 71 | 47 | 34 | - |
| Sex (F:M) | 65 : 22 | 55 : 16 | 37 : 10 | 25 : 9 | 0.926 |
| Surgery type (SG:RYGB:BPD-DS) | - | 46 : 12 : 13 | 30 : 10 : 7 | 23 : 7 : 4 | 0.904 |
| Type 2 Diabetes Mellitus (Y:N) | 21 : 66 | - | - | - | - |
| Age (years) | 44.0 ± 9.2 | 45.0 ± 8.8 | 46.3 ± 9.1 | 49.1 ± 7.5 | 0.034 |
| Weight (kg) | 121.7 ± 14.7 | 94.7 ± 11.6 | 81.3 ± 13.1 | 81.2 ± 14.6 | **< 0.001** |
| BMI (kg/m$^2$) | 43.9 ± 4.2 | 34.2 ± 3.6 | 29.5 ± 4.5 | 29.0 ± 4.0 | **< 0.001** |
| Waist circumference (cm) | 130.2 ± 10.4 | 110.3 ± 10.0 | 100.1 ± 11.9 | 98.6 ± 11.2 | **< 0.001** |
| Hip circumference (cm) | 132.7 ± 10.8 | 114.8 ± 9.5 | 105.6 ± 9.6 | 106.0 ± 10.8 | **< 0.001** |
| Neck circumference (cm) | 41.3 ± 3.5 | 37.0 ± 2.8 | 35.2 ± 3.3 | 34.8 ± 3.2 | **< 0.001** |
| Excess weight loss (%) | - | 46.6 ± 9.7 | 70.8 ± 16.3 | 72.0 ± 15.2 | **< 0.001** |
| Total weight loss (%) | - | 21.7 ± 4.0 | 33.1 ± 7.2 | 33.5 ± 7.4 | **< 0.001** |
| Systolic blood pressure (mmHg) | 134 ± 15 | 121 ± 13 | 119 ± 16 | 116 ± 13 | **< 0.001** |
| Diastolic blood pressure (mmHg) | 80 ± 11 | 74 ± 10 | 72 ± 11 | 69 ± 9 | **< 0.001** |
| Fasting glycemia (mmol/L) | 6.2 ± 1.6 | 5.1 ± 0.9 | 4.8 ± 0.7 | 5.0 ± 0.8 | **< 0.001** |
| Fasting insulin (pmol/L) | 169.5 ± 94.1 | 65.5 ± 40.2 | 46.3 ± 31.4 | 43.7 ± 21.5 | **< 0.001** |
| HOMA-IR index | 8.1 ± 5.4 | 2.6 ± 2.1 | 1.7 ± 1.4 | 1.7 ± 0.9 | **< 0.001** |
| Total cholesterol (mmol/L) | 4.5 ± 0.9 | 4.0 ± 1.1 | 4.2 ± 0.9 | 4.3 ± 0.8 | 0.005 |
| LDL-cholesterol (mmol/L) | 2.6 ± 0.8 | 2.3 ± 1.1 | 2.3 ± 0.8 | 2.3 ± 0.7 | 0.219 |
| HDL-cholesterol (mmol/L) | 1.2 ± 0.3 | 1.2 ± 0.3 | 1.4 ± 0.3 | 1.5 ± 0.3 | **< 0.001** |
| Triglycerides (mmol/L) | 1.63 ± 0.8 | 1.4 ± 0.8 | 1.1 ± 0.5 | 1.0 ± 0.5 | **< 0.001** |

Results are presented as mean ± SD; repeated-measures ANOVA by linear model (Chi-square for categorical variables) comparing baseline, 4 months, 12 months, and 24 months post-surgery sessions. SG, sleeve gastrectomy; RYGB, Roux-en-Y gastric bypass; BPD-DS, biliopancreatic derivation with duodenal switch; BMI, body mass index; HOMA-IR, homeostatic model assessment for insulin resistance; HDL, high-density lipoprotein; LDL, low-density lipoprotein

**Voxel-based morphometry**

Grey matter density across cortical and subcortical regions as measured by voxel-based morphometry (VBM) were used to train the brain age prediction model and assess the relationship between predicted brain age, obesity, and weight loss. Briefly, all T1-weighted structural scans were processed through a standard VBM pipeline using the following steps: 1) image denoising[39]; 2) intensity non-uniformity correction[40]; and 3) image intensity normalization into range (0–100) using histogram matching. Using the ANIMAL software[41], the T1-weighted images were segmented into grey matter, white matter, and cerebrospinal fluid images. VBM analysis was performed using MNI MINC tools to generate grey matter density maps per voxel. Images were then first linearly (using a nine-parameter rigid registration) and then nonlinearly registered to an average brain template (MNI ICBM152-2009c)[42] using MNI MINC tools (http://www.bic.mni.mcgill.ca/ServicesSoftware/MINC) and Advanced Normalization Tools (ANTS) software (http://stnava.github.io/ANTs/), respectively.

**Parcellation used for regularization and exploratory analysis**

For cortical regions, Schaffer functional MRI parcellation at 1000 regions[43] was used to extract mean regional VBM values from each map. For subcortical regions, we used the atlas developed by Xiao et al.[44], including 11 subcortical regions in each hemisphere (http://nist.mni.mcgill.ca/?p=1209).

*Brain age prediction model based on the training dataset*

We used principal component analysis (PCA) as the dimension reduction method to extract VBM-based brain measures for the prediction models. PCA is a singular value decomposition based data factorization method commonly used in brain age prediction studies[45–47]. We used linear regression as the main prediction model to predict the brain age, using VBM-based PCAs as predictive features. 10-fold cross validation was used to ensure generalizability of the model and avoid overfitting. We used the prediction accuracy based on the 10-fold cross validation to assess the performance of the model in the training sample (i.e. CamCAN dataset) and select the optimal number of features included in the final model. Root-mean-squared error (RMSE) was used as the natural cost function for the linear regression model. Furthermore, all the analyses were repeated using the linear regression model with least absolute shrinkage and selection operator (LASSO) regularization to ensure robustness based on regional VBM values. The LASSO algorithm was used as implemented in *fitrlinear* and *lasso* functions in MATLAB 2021a.

*Delta age and adjusted delta age*

After predicting the brain age as explained above, we calculated the difference between the predicted brain age and the chronological age (i.e. delta age). Delta age can be used as an estimate of brain-related health, since it measures the discrepancy between brain age based on the grey matter density and the expected brain age (chronological age).

(1) $Y = X * \beta_1 - \delta_1 \rightarrow \delta_1 = X * \beta_1 - Y$

In formula (1), $\delta_1$, $X$, and $Y$ indicate delta age, VBM-based principal components, and chronological age, respectively. Due to its definition, delta age is correlated with the chronological age of the participants, which by nature will act as a confounder in the analysis and will make it difficult to distinguish the effect of chronological age and the additional biological delta age. There have been several different adjustments proposed to correct for this effect[46,48–50], here we use the Smith et al. 2019[46] definition, regressing out the portion of delta age explained by chronological age (formula (2)). We call the residual, the adjusted delta age.

(2) $\delta_2 = \delta_1 - Y * \beta_2$

Adjusted delta age has been used as the main measure of interest in the manuscript.

**Statistical Analyses**

*Independent sample analysis*

Based on the trained models, brain age was predicted for the participants from the HCP dataset using the full sample model trained on the CamCAN dataset. We then calculated the adjusted delta age for the test sample based on age matched training sample and test set and performed paired t-tests to compare predicted brain age values for individuals with obesity versus individuals with normal weight.

*Bariatric surgery-induced weight loss and brain age*

Using the full-sample model trained on CamCAN dataset, brain age was predicted for the bariatric study participants prior to surgery (baseline), as well as 4 months, 12 months, and 24 months post-surgery. As explained above, we calculated the delta age and adjusted delta age with bias correction for the bariatric sample. We used a mixed effects model (model I) to examine the effect of visits on adjusted delta age, controlling for sex, age, BMI, and diabetic status at baseline as well as the surgery type and with subjects as categorical random effects. One tailed unpaired t-tests were used as post hoc tests to assess whether adjusted delta age significantly decreased post-surgery compared to baseline.

**model I**: Adjusted Delta age ~ Visit + Age + Sex + $BMI_{baseline}$ + Diabetic $status_{baseline}$ + Surgery type + (1|Subject)

*Associations between changes in cardiometabolic risk factors following surgery and delta age*

To examine the association between the changes in cardiometabolic variables and delta brain age, we used a mixed effects model (model II) with adjusted delta age as the dependent variable and the cardiometabolic variables as the fixed effects of interest, while controlling for BMI and diabetic status at baseline, age, sex, and surgery type, and including subject as a categorical random effect:

**model II:** Adjusted Delta age ~ Cardiometabolic variable + Age + Sex + $BMI_{baseline}$ + Diabetic $status_{baseline}$ + Surgery type + (1|Subject)

Due to the limited sample size and the multi-collinearity, the effect of each cardiometabolic variable was examined separately. Metabolic variables in this analysis included BMI, systolic/diastolic blood pressure, plasmatic levels of triglycerides, LDL-cholesterol, HDL-cholesterol and glucose, as well as the HOMA-IR. The results were corrected for multiple comparisons using Bonferroni method with a significance threshold of 0.05.

*Exploratory regional analysis*

In order to investigate the variability in delta age across brain regions, we used a parcellation-based approach to measure grey matter VBM values across the brain. We used these regional VBM values as predictors and performed a similar prediction-based analysis with each single brain region. We then calculated the adjusted delta age at baseline and across the 4, 12, and 24 months after surgery and compared the adjusted delta age post- and pre-surgery using paired t-tests. The results were corrected for multiple comparisons using False Discovery Rate (FDR) controlling method with a significance threshold of 0.05. While single-region prediction is more interpretable and targets regional specificity of the effect, the prediction accuracy is much lower at this level and the analysis should be treated as exploratory and the results should be interpreted with caution.

All prediction and statistical analyses were performed using MATLAB 2021a.

**Results**

*Brain Age Prediction Model*

The linear regression model with principal components of the grey matter VBM was able to predict chronological age, yielding a cross-validated RMSE value of 8.8 (52% increase compared to mean) and correlation of r=0.90 (p<0.0001). We used RMSE as the cost function for age prediction and evaluated the model based on the number of principal components included (for the training dataset), obtained through 10-fold cross validation. We chose the number of principal components resulting in the minimum RMSE (N=45) as the optimal number of the principal components, and therefore the predictive models are based on 45 VBM-based principal components as the predictive features. These principal components are projected to the out of sample datasets (i.e. HCP and Bariatric surgery datasets) to calculate the features for the analyses. While the results obtained from regional analysis using LASSO was similar, the overall performance of the model was considerably worse (cross-validated RMSE 9.7 years) and therefore all the results are reported based on the voxel-based models.

*Obesity and Brain Age*

**Figure 1** shows boxplots of adjusted delta age for matched individuals with obesity and individuals with normal weight from the HCP dataset. Participants with obesity had significantly higher delta age values reflecting poorer brain health than the matched participants with normal weight (T-value = 7.08, p-value < 0.0001).

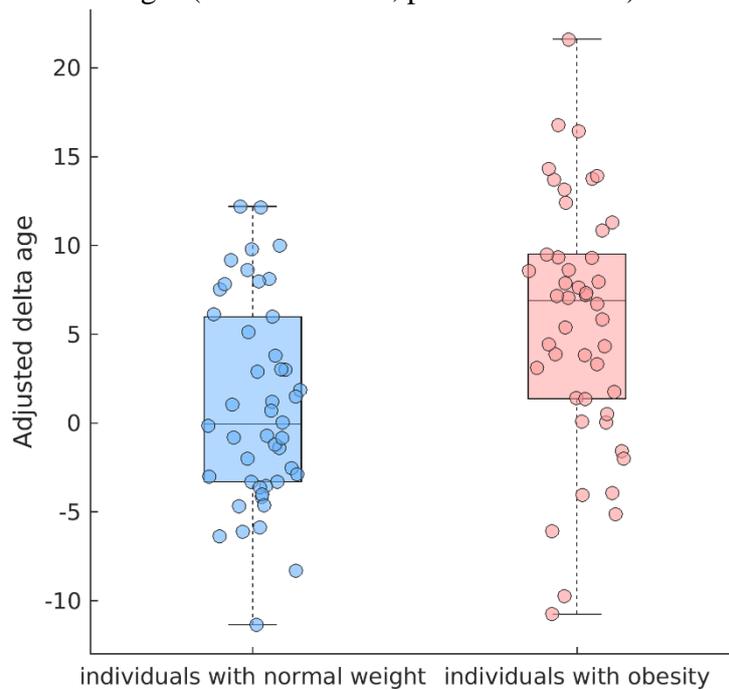

**Figure 1.** Adjusted predicted age for individuals with obesity and normal weight in the HCP dataset. There is a significant difference between the adjusted brain age between the two groups (T-value = 7.08, p-value < 0.0001).

*Weight Loss and Brain Age*

**Figure 2** shows boxplots of adjusted delta age predicted based on the MRIs acquired at baseline (the visit prior to surgery) as well as the three visits (4 months, 12 months, and 24 months) post-surgery. We found a significant effect of visits using a mixed effects model controlling for sex, age, and BMI at baseline as well as the surgery type (model I). The post hoc unpaired t-tests showed significant decreases in delta age 12-month post-surgery (T-value= -3.76, p-value < 0.0005) and 24-month post-surgery visits compared to baseline (T-value = -6.40, p-value < 0.0005), but not at 4-month post-surgery (T-value = -0.25, p-value = 0.79) compared to baseline.

Furthermore, as a post hoc confirmation, we also compared the adjusted delta age values between the training dataset with cross validation (i.e. CamCAN dataset with healthy participants) and the pre- and post-surgery visits. We expected to observe higher adjusted delta age in bariatric surgery samples in early visits compared to the participants from CamCAN and similar delta age at later visits when participants return to normal weight and the sample is comparable to the training sample. The delta age at baseline, 4 month post-surgery, and 12 month post-surgery was higher compared to the delta age of training dataset (all p-values < 0.009), while the delta age at 24 month post-surgery was not significantly different from individuals of the training dataset with normal weight (p-value = 0.80).

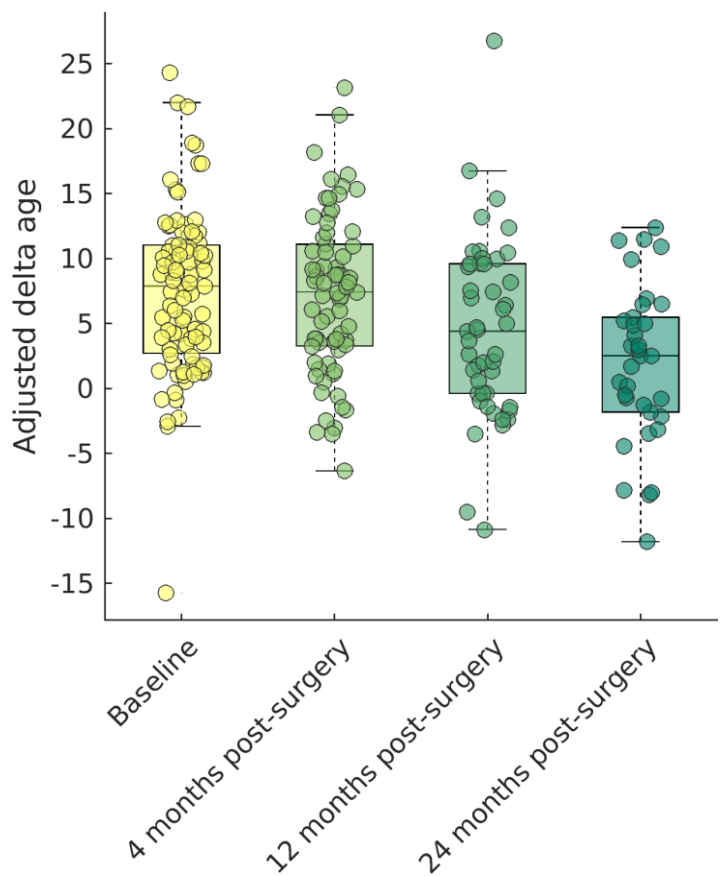

**Figure 2.** Adjusted predicted age for individuals in the bariatric surgery dataset at baseline (pre-surgery), 4, 12, and 24 months post-surgery.

*Associations between Delta Brain Age and Cardiometabolic Risk Factors*

We used mixed effect models (model II) to examine the associations between the delta brain age and adiposity/cardiometabolic factors including BMI, systolic/diastolic blood pressure, plasmatic levels of triglycerides, LDL-cholesterol, HDL-cholesterol, and glucose as well as the HOMA-IR. We found a significant association between adjusted delta age and BMI, systolic/diastolic blood pressure, and HOMA-IR (T-value$_{BMI}$=3.59, T-value$_{SBP}$=4.26, T-value$_{DBP}$=3.67, T-value$_{HOMA-IR}$=2.81 respectively, all Bonferroni corrected p-values < 0.05) where higher BMI, blood pressure, and HOMA-IR was related to higher delta age. We did not find any significant association between adjusted delta age and the rest of cardiometabolic variables.

*Exploratory Regional Analysis*

While the predictive model based on PCA provides a global perspective of brain age and health, we cannot specify the brain regions contributing to overall differences. In order to identify variability across brain regions, we used a parcellation-based approach with VBM values as predictive features. We repeated the age prediction and brain age calculation using one brain region in each model. While as expected, the model performance is much lower (as measured by RMSE) than whole brain analysis, this exploratory analysis provides some regional insights. **Figure 3** shows the brain regions with significantly lower adjusted delta age at 24 months post-surgery compared to baseline mainly in somatomotor, visual, and ventral attention networks. We did not find any significant differences at single region level at 4 months and 12 months post-surgery.

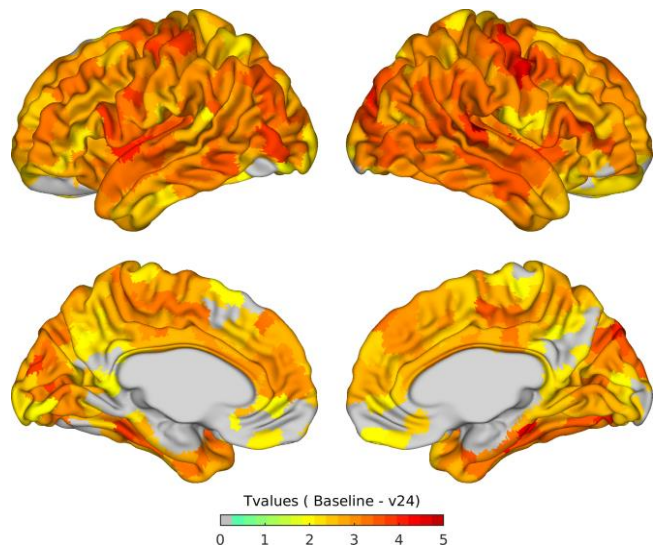

**Figure 3.** T-values comparing the adjusted delta age between baseline and 24 months post-surgery. Brain age prediction model was estimated based on each brain region separately in the CamCAN dataset and was applied to the bariatric surgery data to calculate the adjusted delta age across visits. The figure shows the T-values for the significant differences (after FDR correction) between adjusted delta age at 24 months post-surgery compared to baseline using an unpaired t-test. We found no significant differences between the adjusted delta age at 4 and 12 months post-surgery compared to baseline.

**Discussion**

In this study, we assessed the impact of weight loss and cardiometabolic improvement following bariatric surgery on brain health, as measured by change in brain age estimated based on VBM measurements. Using an independent dataset from HCP, we found a higher brain age for individuals with obesity compared to individuals with normal weight. Our results also showed significant improvement in brain health, indicated by a decrease of 2.9 and 5.6 years in adjusted delta age at 12 and 24 months following bariatric surgery as compared to baseline. Adjusted delta age at 24 months post-surgery was not different from individuals of the training dataset with normal weight in the training dataset (CamCAN). While the overall effect seemed to be driven by a global change across all brain regions and not from a specific region, our exploratory analysis showed lower delta age in certain brain regions (mainly in somatomotor, visual, and ventral attention networks) at 24-month. This reduced age was also associated with post-surgery improvements in BMI, systolic/diastolic blood pressure, and HOMA-IR.

Our results based on the participants from the HCP study are in line with previous studies in the literature indicating associations between obesity and poorer brain health as reflected in higher delta age in individuals with obesity compared to those with normal weight. In 234 participants, Kolenic et al.[1] found a significant and positive association between BMI and brain age scores. Smith et al. also reported a strong relationship between BMI and brain age assessed with multimodal imaging for 19,038 participants from the UK Biobank[46]. Waist-to-hip ratio was also associated with delta age calculated based on DTI data[28]. In a cross-sectional study, overweight and obesity was associated with an estimated 10-year increase in brain age based on white matter volume[2]. These brain age differences are likely due to grey matter and white matter atrophy in individuals with obesity[51]. It has been suggested that part of these structural brain alterations observed in obesity might be attributed to the cardiometabolic burden that abdominal adiposity entails[51]. For instance, chronic low-grade inflammation related to abdominal obesity has been associated with disruptions in grey and white matter integrity as well as cerebrovascular disease[27,52,53]. A recent large-scale study from the UK Biobank cohort (~20 000 participants) found that inflammation, hypertension, and type 2 diabetes are associated with brain small vessel disease (as measured by volume of white matter hyperintensities), which is in turn related to changes in brain cortical and subcortical morphology. These changes also appear to be linked to poorer cognitive performance[24]. The obesity-related cardiometabolic alterations might lead to negative effects on brain health by disrupting cerebral blood flow and its supply[51]. A few recent large-scale studies have also reported positive associations between delta age and cardiometabolic risk factors, including blood pressure, smoking, alcohol intake, and stroke risk[28,29]. These cardiometabolic risk factors can contribute to the development of atherosclerosis, leading to cerebral ischemia, microstructural damage, and cognitive impairment[54]. Taken together, these findings are in favour of the hypothesis that obesity-related cardiometabolic alterations might lead to cerebral grey and white matter abnormalities and poorer brain health.

Comparing delta age values prior to and after bariatric surgery showed a steady improvement in brain health, as reflected in significant decrease in delta age values at 12 months and 24 months post-surgery. These results are in line with previous MRI studies from our group[19–21] and others[55–57] showing widespread increases in white matter and grey matter densities as well as resting neural activity (as measured by fractional-amplitude of low frequency fluctuations) following bariatric surgery, suggesting a global effect of surgery on brain status. These grey matter density increases were more pronounced and widespread 12 months post-surgery compared to 4 months post-surgery[21], which could explain why the delta age values at 4 months post-surgery were similar to baseline levels. The amount of weight loss and the improvement in cardiometabolic factors were also less significant 4 months post-surgery compared to 12- and 24-months post-surgery. The improvement in brain age following the surgery is also in line with several studies showing a delay or a slowing of aging processes with caloric restriction and/or weight loss[58,59]. Comparing the training dataset from CamCAN with pre- and post-surgery visits, we also found that delta age values at 24 months post-surgery were similar to delta age values from individuals with normal weight, suggesting that bariatric surgery and the consequent weight loss could improve brain health and reverse the observed accelerated aging signature in the brain.

Decrease in delta age values following bariatric surgery was significantly associated with the degree of weight loss and concomitant improvement in cardiometabolic factors, more specifically blood pressure and insulin resistance. However, no significant association was observed with changes in lipid profile (LDL-cholesterol, HDL-cholesterol, or triglycerides).

This is consistent with the study from Kolenic et al. that did not find significant associations between delta age and lipid profiles[1]. Our results are also consistent with our previous findings showing that the increase in grey matter and white matter densities after bariatric surgery are correlated with weight loss and improvement of the metabolic/inflammatory profiles[21]. Together, these findings support the idea that the improvement in brain integrity and health, as indexed by delta age, could be a consequence of a better cerebral blood flow and improved insulin sensitivity after bariatric surgery-induced weight loss. Recent meta-analyses provide strong evidence that bariatric surgery is associated with improvement in subclinical atherosclerosis, artery endothelial function, and resolution of hypertension[60,61]. These cardiometabolic/vascular changes observed following bariatric surgery may lead to improved angiogenesis and changes in synaptic connectivity, dendritic branching, axon sprouting, and glial cells, which could influence grey matter density as well as brain health[62,63]. However, further studies are needed to better understand the underlying mechanisms through which the cardiometabolic improvements following bariatric surgery influence brain integrity and brain health.

Some limitations should be addressed. We did not control for some factors associated with brain aging such as smoking, alcohol consumption, education, and physical activity levels[64]. Despite previous reports of sex-specific differences in brain aging and cardiometabolic factors[30], due to sample size limitations, we were unable to explore changes in men and women separately. Indeed, the bariatric population normally includes a lower proportion (approximately 25%) of men. Thus, we recruited a similar proportion of men in the current study. Larger sample sizes are needed to further investigate sex-specific differences in brain delta age and their associations with weight-loss and cardiometabolic factors. Furthermore, we used brain age as proxy measure for improved brain health since we didn't have access to the direct measures of brain health (e.g. cerebrospinal fluid measures). However these measures have been reported to be associated with brain age[7,17].

The brain age prediction model had an RMSE of 8.7 (for 18.61 years standard deviation of age), comparable to what has been previously observed in the literature, and using the same dataset[2,4]. A 10-fold cross validation scheme was used to ensure that the results were not impacted by leakage[65]. The same training dataset used here has been used in the study by Ronan et al.[2] However, to assess the impact of obesity on delta age, we used an age-matched independent sample from HCP, to ensure that the findings are not biased by the fact that older people in general tend to be more overweight and obese.

In conclusion, our study revealed significant obesity-related differences in brain health (as measured by delta age) between individuals with obesity and those with normal weight, as well as a marked improvement in brain age following bariatric surgery. These results suggest that obesity-related brain health abnormalities might be reversed by means of significant and sustained weight-loss, along with widespread improvements in cardiometabolic alterations.

**Acknowledgements**

We would like to acknowledge the contribution of surgeons, nurses, the medical team of the bariatric surgery program at IUCPQ, MRI technicians, Xavier Moreel, Coordinator of the Plateforme d'imagerie avancée at IUCPQ, and Guillaume Gilbert, Engineer, Phillips as well as the collaboration of participants. Data collection for training dataset was provided by the Cambridge Centre for Ageing and Neuroscience (CamCAN). CamCAN funding was provided by the UK Biotechnology and Biological Sciences Research Council (grant number BB/H008217/1), together with support from the UK Medical Research Council and University of Cambridge, UK. HCP data was obtained from the Human Connectome Project, WU-Minn Consortium (Principal Investigators: David Van Essen and Kamil Ugurbil; 1U54MH091657) funded by the 16 NIH Institutes and Centers that support the NIH Blueprint for Neuroscience Research; and by the McDonnell Center for Systems Neuroscience at Washington University. Authors thank Compute Canada (https://www.computecanada.ca/home) for the usage of the computing resources in the current work.

**Funding**

This study is supported by a Team grant from the Canadian Institutes of Health Research (CIHR) on bariatric care (TB2-138776) and an Investigator-initiated study grant from Johnson & Johnson Medical Companies (Grant ETH-14-610). Funding sources for the trial had no role in the design, conduct or management of the study, in data collection, analysis or interpretation of data, or in the preparation of the present manuscript and decision to publish. This research was undertaken thanks in part to funding from the Canada First Research Excellence Fund, awarded to McGill University for the Healthy Brains, Healthy Lives (HBHL) initiative. MD is supported by the Alzheimer Society Research Program (ASRP) postdoctoral award. The Co-investigators and collaborators of the REMISSION study are (alphabetical order): Bégin C, Biertho L, Bouvier M, Biron S, Cani P, Carpentier A, Dagher A, Dubé F, Fergusson A, Fulton S, Hould FS, Julien F, Kieffer T, Laferrère B, Lafortune A, Lebel S, Lescelleur O, Levy E, Marette A, Marceau S, Michaud A, Picard F, Poirier P, Richard D, Schertzer J, Tchernof A, Vohl MC.

**Disclosure statement**

A. T. and L. B. are recipients of research grant support from Johnson & Johnson Medical Companies and Medtronic for studies on bariatric surgery and the Research Chair in Bariatric and Metabolic Surgery at IUCPQ and Laval University. AT has received consulting fees from Bausch Health, Novo Nordisk and acts as a consultant for Biotwin. No author declared a conflict of interest relevant to the content of the manuscript.